# Control of OSIRIS-REx OTES Observations using OCAMS TAG Images

Kris J. Becker[1] and Kenneth L. Edmundson[2]

[1] Lunar & Planetary Laboratory, University of Arizona, Tucson, AZ 85721, USA
[2] Edmundson Photogrammetry Consulting LLC, Flagstaff, AZ 86004, USA

## Abstract

As the OSIRIS-REx spacecraft descended toward the asteroid Bennu to collect a sample from the surface in the touch-and-go (TAG) procedure, many of the instruments were actively collecting observation data. We applied the process of photogrammetric control to accurately determine the position and attitude of 190 OCAMS MapCam and SamCam descent images at the time of exposure. The average image pixel resolution is 10cm (median is 7cm). The images were controlled to ground using simulated images generated from high resolution (5cm, 44cm and 88cm ground sample distance) shape models of Bennu. After least-squares adjustment, the root mean square (rms) of all image measurement residuals was 0.16 pixels. These results were applied to 581 OTES observations by interpolation over the updated ephemeris of the OCAMS MapCam and SamCam instruments using frame transformations from OCAMS to the OTES frame. Then, the surface intercept of the OTES field of view was recomputed by ray tracing the adjusted boresight look direction onto the 44cm shape model. The average of the adjusted OTES boresight surface intercepts differed from the a priori locations on the 88cm shape model by ~37cm with an uncertainty less than 5cm.

## Introduction

The OSIRIS-REx (OREX) touch-and-go (TAG) operation to collect a sample from the surface of asteroid 101955 Bennu was successfully executed on October 20, 2020. The spacecraft acquired extensive data with many of its instruments in the science payload (Figure 1) throughout the TAG event. Here we describe the registration (or control) of OSIRIS-REx Thermal Emission Spectrometer (OTES, Christensen et al., 2018) TAG observations to the surface of Bennu through the rigorous photogrammetric control of TAG images collected by the MapCam and SamCam cameras of the OSIRIS-REx Camera Suite (OCAMS, Rizk et al., 2018).

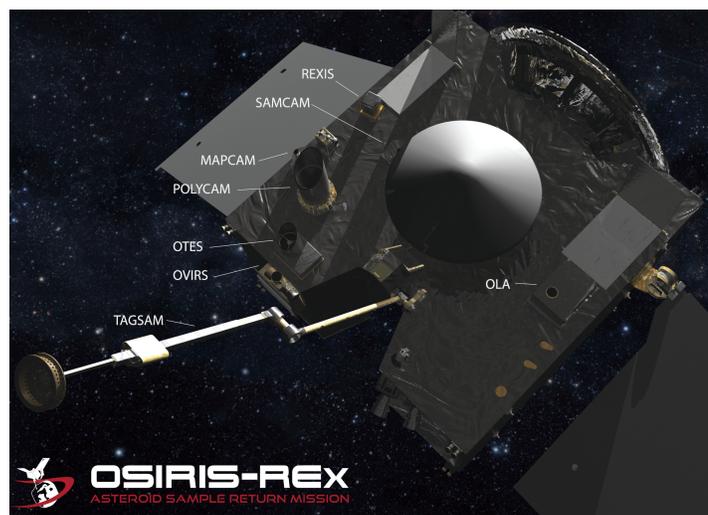

*Figure 1. OSIRIS-REx Science Payload Deck showing relative instrument positions.*

OTES is a point (or spot) spectrometer. Surface features captured in an OTES observation can be established by accurately correlating the spot center with overlapping MapCam and SamCam images. This in turn can be compared to the spot center as determined from thermal model offset analysis. Our goal was to establish spot center positions to within 10 cm.

While the a priori spacecraft ephemeris and pointing kernels determined by the OSIRIS-REx team are of high quality, some degree of uncertainty is inevitable. This may result from, for example, differences in shape model and coordinate systems, uncertainties in NAIF (NASA's Navigation and Information Facility; Acton, 1996) ephemeris kernels, or other dynamic physical properties of the TAG environment. To minimize errors and achieve our accuracy requirement, images are controlled using the least-squares bundle adjustment (Brown, 1958). Input to the bundle adjustment includes two-dimensional point feature measurements from multiple overlapping images and a priori values for surface point coordinates and image position and orientation. The adjustment simultaneously solves for refined values of these parameters.

As OTES is a point spectrometer, it's surface observations are typically too small to control via photogrammetry. Here, the photogrammetric control process was first performed on MapCam and SamCam TAG images, yielding improved spacecraft position and camera pointing parameters for each. Control was accomplished with a modified version of the Integrated Software for Imagers and Spectrometers (ISIS) planetary cartography package developed and maintained by the OSIRIS-REx Image Processing Working Group (IPWG) (DellaGiustina et al., 2018; Sides et al., 2017). Version 21 (V021) of the high-resolution 5cm regional and 44cm global average ground sample distance (GSD) shape models of Bennu were used for ground registration. With these results, improved spacecraft position and pointing values at the time of each OTES observation were then determined through interpolation. Finally, the spot center surface location for each OTES observation was computed by intersecting the adjusted boresight look vector with the Bennu shape models via ray tracing.

Below we first describe the photogrammetric control of the OCAMS TAG images. This is followed by a description of the OTES TAG observations and the interpolation methods used to improve their position and pointing.

## Control of OCAMS TAG Images

There were 247 OCAMS images acquired before and after TAG. We controlled 190 of the descent images obtained prior to TAG (78 MapCam and 112 SamCam) as the surface was not visible due to the resulting debris afterwards. The control pipeline closely follows what we have established for all other OCAMS control projects (e.g. Bennett et al., 2020; Edmundson et al. 2020).

### Kernels

Kernels used for processing of the OCAMS images are shown in Table 1. The IPWG version of ISIS can utilize a prioritized list of shape models. This has proven crucial when multiple high precision shape models are required to process images contained in a control network or mosaic.

Table 1. Primary kernels used for processing OCAMS images.

| | |
|---|---|
| NaifFrameCode | -64362 |
| NaifSpkCode | -64362 |
| LeapSecond | $base/kernels/lsk/naif0012.tls |
| TargetAttitudeShape | $osirisrex/kernels/pck/bennu_v17.tpc |
| TargetPosition | $osirisrex/kernels/tspk/de424.bsp |
| | $osirisrex/kernels/tspk/bennu_refdrmc_v1.bsp |
| | $osirisrex/kernels/tspk/sb-101955-76.bsp |
| | $osirisrex/kernels/pck/pck00010.tpc |
| InstrumentPointing | $osirisrex/kernels/ck/orx_sc_rel_201019_201025_v00.bc |
| | $osirisrex/kernels/fk/orx_v14.tf |
| Instrument | $osirisrex/kernels/ik/orx_ocams_v07.ti |
| SpacecraftClock | $osirisrex/kernels/sclk/ORX_SCLKSCET.00083.tsc |
| InstrumentPosition | $osirisrex/kernels/spk/orx_200827_201020_201020_od294-N_tag_v1.bsp |
| | $osirisrex/kernels/spk/orx_201020_201115_201109_od294-N_postbackaway_v1.bsp |
| InstrumentAddendum | orex_ocams_addendum_v12.ti |
| ShapeModel | $osirisrex/kernels/dsk/l_00050mm_alt_ptm_5595n04217_v021.obj |
| | $osirisrex/kernels/dsk/bennu_g_00400mm_alt_ptm_0000n00000_v021.bds |
| InstrumentPositionQuality | Unknown |
| InstrumentPointingQuality | Reconstructed |
| Extra | $osirisrex/kernels/spk/orx_struct_v04.bsp |
| RayTraceEngine | Bullet |

## Control Point Measurement

The photogrammetric control process improves the registration between the images themselves and between the images and an associated shape model. Overlapping images are registered to one other by measuring common features known as tie points (TPs). Images are registered to the ground by measuring features common between them and an existing map, mosaic, or shape model. These are called ground control points (GCPs).

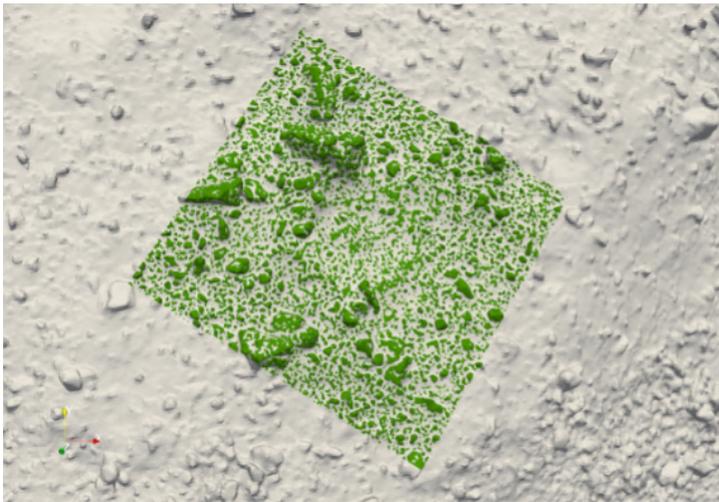

Figure 2. Bennu V021 5cm (green) and 44cm shape models superimposed show very good alignment between surface features.

Tie point image measurements are acquired in ISIS by automatic interest point detection and image matching with the *findfeatures* application. The *pointreg* application is then used to refine these measurements to sub-pixel accuracy (Becker et al., 2017).

To generate GCPs we first utilized the Bennu V021 5cm and 44cm GSD shape models (Figure 2) to produce a simulated image corresponding to each of the 190 OCAMS images (e.g. Golish et al., 2017; Becker et al., 2020). The 44cm shape model was needed to complete coverage where the 5cm did not cover all of the descent images. It is generally easier to identify features in the simulated images than directly in the shape models as they are more realistic in appearance due to the accurate representation of the light source and the presence of shadows. The simulated images and OCAMS images with common ground coverage were then feature-matched to produce GCPs. Common GCPs were merged and refined to subpixel accuracy. The final control network contains 3,335 points and 55,228 image measurements.

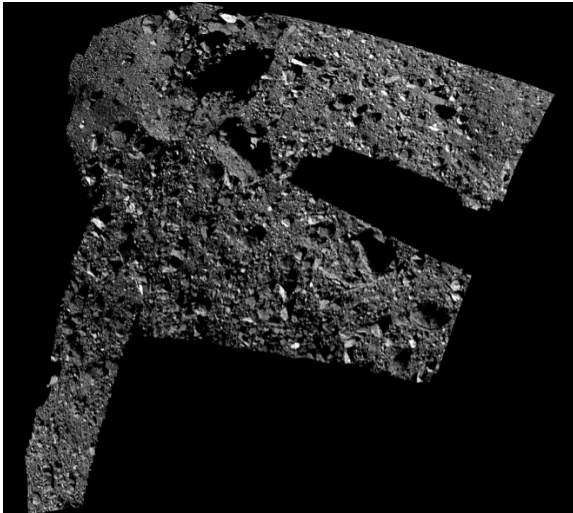
*Figure 3. Mosaic of OCAMS MapCam and SamCam TAG images after bundle adjustment.*

Prior to automated control processing, a mask was created for the SamCam images to remove the TAGSAM arm from some of the images. This was required as it would adversely impact image matching and registration by creating control points/measures on the TAGSAM assembly itself. Consequently, there is a swath of missing data in the final controlled mosaic produced after bundle adjustment (Figure 3).

## OCAMS Bundle Adjustment Results

The final control network consisted of 190 images, 3,309 TPs, 26 GCPs, and 55,228 image measurements (Figure 4, Figure 5). Bundle adjustments were performed with the ISIS application *jigsaw* (Edmundson et al., 2012). Initial image position and pointing parameters (jointly referred to as exterior orientation) were obtained from reconstructed NAIF spacecraft position and camera pointing kernels. Initial TP coordinates were determined by first projecting all image measurements through the appropriate sensor model to obtain the intersection with the shape model. For each TP, all measurement intersection points were averaged to obtain coordinates for input to the bundle adjustment. TP latitude and longitude coordinates were allowed to freely adjust. All TP radius coordinates were assigned an a priori uncertainty of 10m. GCP coordinates were held fixed (i.e. not permitted to adjust) to their coordinates obtained from the simulated images/shape models. A priori image position and pointing uncertainties were 15m and 0.1° respectively.

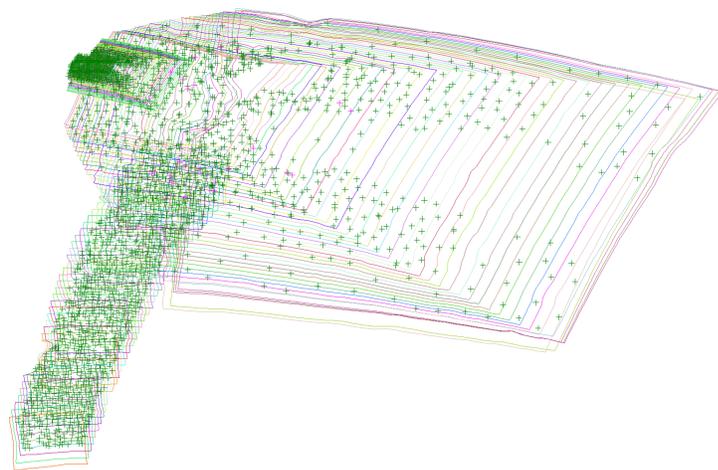
*Figure 4. TAG descent control network image measurement distribution. 55,228 measurements overlaid on 190 MapCam and SamCam footprints.*

The bundle adjustment converged in six iterations. The root mean square (rms) of all image measurement residuals was 0.16 pixels. The rms of TP image measure residuals was 0.13 pixels. The residual rms for GCPs was ten times that of the TPs at 1.29 pixels. This discrepancy is a result of holding the GCP coordinates fixed in the adjustment at a priori values that have some uncertainty.

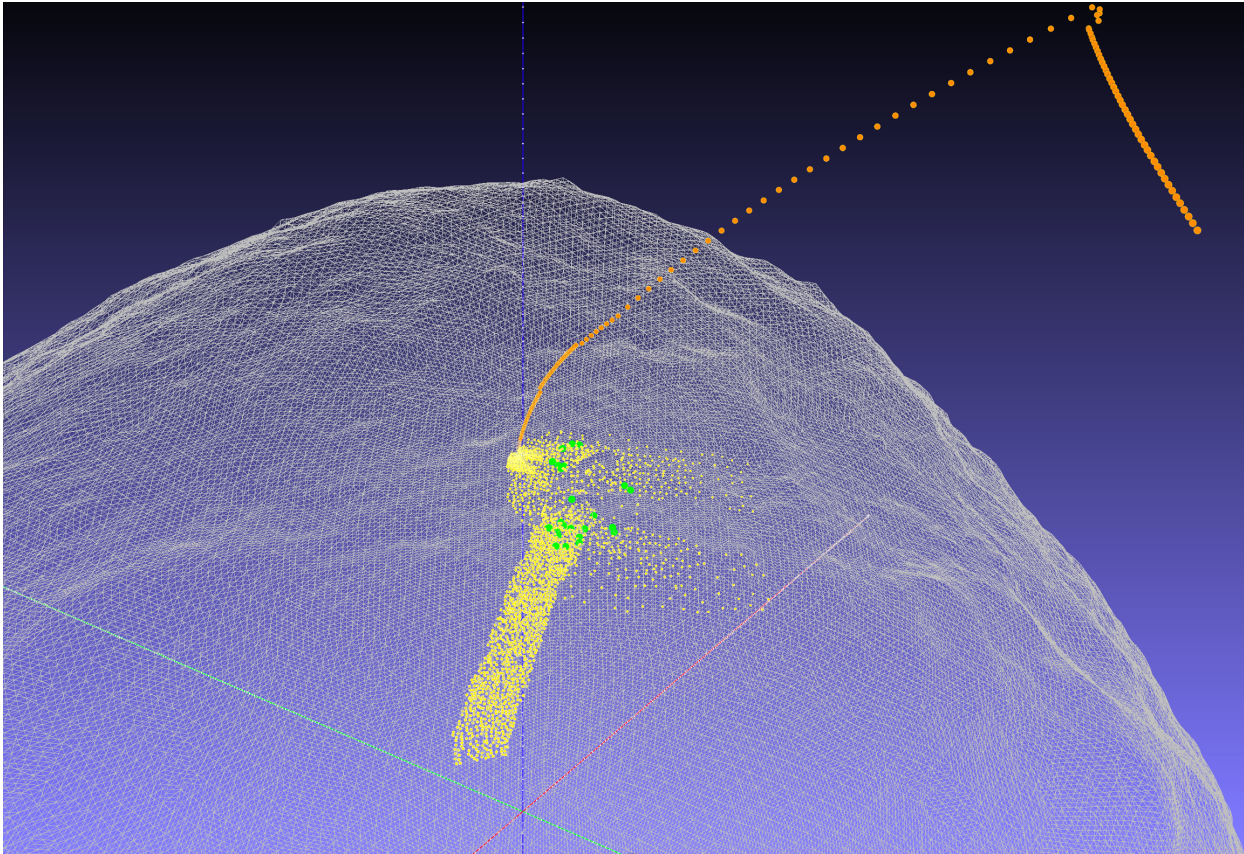

*Figure 5. 3D view of ground point distribution and TAG spacecraft positions. Tie points (TPs) are shown in yellow, ground control points (GCPs) in green, and spacecraft positions in orange. Points are shown relative to a mesh approximating Bennu's surface.*

## Anatomy of ISIS Cube Geometry Data

The ISIS cube stores ephemeris data In *Table*s, a generic data structure that contains rows and columns of binary values. There are two tables of position data – InstrumentPosition and SunPosition and two tables of rotation data – InstrumentPointing and BodyRotation. These four ephemeris data components provide all the necessary elements to compute a complete set of cartographic and high-precision geometric data in the ISIS Camera model system.

The ISIS *spiceinit* application initializes these data tables with NAIF SPICE kernels based on the observation time of the acquired image data. For the OCAMS instruments, which are framing cameras, only one row of ephemeris data is contained in the position and rotation data at the center of the exposure time.

In each row of the position table there are seven double precision values. The position table data is comprised of a 3-vector of XYZ position (km) data, an optional 3-vector of velocity (km/s) of the instrument or sun, and a scalar value of the ephemeris time (seconds) at the center of the observation.

Each row of the rotation tables contains eight double precision values. The first four values in the row are a NAIF engineering-type quaternion of four values (WXYZ) (radians), three values of angular velocity (km/s), and a single scalar value of the ephemeris time (seconds) at the center of the observation.

The InstrumentPointing and InstrumentPosition table contents are updated from the bundle adjustment solution. The contents of these tables can then be used to calculate the position and orientation of any of the other OREX instruments using (typically) fixed frame transformations provided in NAIF SPICE kernels.

## OTES TAG Observations

OREX TAG observations used were acquired from 2020-10-20T21:27:02.137 to 2020-10-20T21:50:52.121 UTC, including 581 OTES surface observations. Initial latitude and longitude locations of the boresight surface intercept were computed using a priori spacecraft and planet ephemerides with the Bennu V020 88cm global shape model as defined by the OREX Science Processing Operations Center (SPOC) metakernel for TAG. The full set of NAIF SPICE kernels in the SPOC metakernel used to compute the a priori OTES locations is given in Table 2.

*Table 2. NAIF Spice kernels used to compute a priori OTES surface locations.*

```
Created by: digestotes2
Coverage start: Tue Oct 13 2020 10:42:08 GMT-0700 (MST)
Coverage end: Tue Oct 27 2020 10:45:26 GMT-0700 (MST)
Type: all
Description: SPOCFLIGHT system metakernel
lsk/naif0012.tls
sclk/ORX_SCLKSCET.00069.tsc
fk/orx_v14.tf
ik/orx_lidar_v00.ti
ik/orx_navcam_v02.ti
ik/orx_ocams_v07.ti
ik/orx_ola_v01.ti
ik/orx_otes_v00.ti
ik/orx_ovirs_v00.ti
ik/orx_rexis_v01.ti
ik/orx_stowcam_v00.ti
ik/orx_struct_v00.ti
pck/pck00010.tpc
pck/bennu_v17.tpc
ck/orx_sc_rel_201012_201018_v01.bc
ck/orx_sc_rel_201019_201025_v01.bc
ck/orx_sc_rel_201026_201101_v02.bc
spk/de424.bsp
spk/bennu_refdrmc_v1.bsp
spk/orx_160909_231024_refod009_v2.bsp
spk/orx_171006_231024_171005_refod027_v1.bsp
spk/orx_200827_201103_201019_od292-N-T1D-L-T1R1_v1.bsp
spk/orx_200827_201020_201020_od294-N_tag_v1.bsp
spk/orx_201020_210524_210103_od297-N-PTO1-F_v1.bsp
spk/orx_struct_v04.bsp
dsk/g_00880mm_alt_obj_0000n00000_v020.bds
dsk/earth_g_0158km_esp_dsk_0000n00000_v101.bds
ck/orx_struct_mapcam_v01.bc
ck/orx_struct_polycam_v01.bc
```

OTES spectral data and other information were retrieved from the SPOC using the query seen in Figure 6. The "otes_hk.et" time value from each OTES observation was used to determine ephemeris data from the OCAMS control interpolated ephemeris data. NAIF SPICE frames, instrument, sclk, spk and ck kernel data were used to calculate frame transforms from the OREX spacecraft frame (ORX_SPACECRAFT) in the

```
curl --request POST \
--url $OTES_DATABASE_URL \
--header 'content-type: application/json' \
-b ~/cookiefile \
--data '{
  "columns": [
        "otes_sci_level2.filename",
        "otes_obs_seq.file_name",
        "otes_obs_seq.target",
        "otes_hk.boresight_intersection_x",
        "otes_hk.boresight_intersection_y",
        "otes_hk.boresight_intersection_z",
        "otes_hk.boresight_direction_x",
        "otes_hk.boresight_direction_y",
        "otes_hk.boresight_direction_z",
        "otes_hk.bore_flag",
        "otes_hk.range",
        "otes_hk.seconds_raw",
        "otes_hk.pm_rot_phase01",
        "otes_sci_temp_spot.kinetic_temp",
        "otes_sci_emiss_spot.emissivity",
        "otes_sci_emiss_spot.xaxis",
        "otes_l2_spectrum_record.ick",
        "otes_hk.otes_look_fk",
        "otes_hk.spkname",
        "otes_hk.ckname_64000",
        "otes_hk.resolution",
        "otes_hk.sclk_string",
        "otes_hk.mid_obs",
        "otes_hk.latitude",
        "otes_hk.longitude",
        "otes_hk.fov_diameter_distance",
        "otes_hk.meta_ker",
        "otes_hk.lonlat",
        "otes_hk.utc",
        "otes_hk.et"
  ],
  "time_utc": {
    "start": "2020-10-20 21:27:02.137Z",
    "end": "2020-10-20 21:50:52.121Z"
  },
  "format": "csv",
  "where": "otes_hk.target_type = '\''T'\''
   and otes_hk.fov_fill_flag = 1
   and otes_l2_spectrum_record.quality <= 2"
}'
```

*Figure 6. Query to retrieve OTES spectral data and additional information.*

J2000 coordinate system to precise OTES frames (ORX_OTES). The boresight and position of the OTES was computed and converted to Bennu body-fixed coordinates using the controlled interpolation models. These positions serve as observer and look direction vector to ray trace onto the Bennu shape models to determine updated latitude and longitude coordinates of the intersection. Other cartographic and geometric (e.g., photometric angles) data can be computed from these results for every OTES observation. In fact, this technique can be applied to any OREX instrument observations acquired during the OCAMS image control timeframe.

## Control Process for OTES Using OCAMS Images

The process to adjust OTES exterior orientation uses the updated control results of the OCAMS descent images during the same time frame as the OTES acquired its observations. There were more OTES observations than OCAMS images. Moreover, many were not acquired at exactly the same time as an OCAMS image. Therefore, the position and orientation of OTES images were determined by interpolation with the OCAMS data.

The 190 OCAMS images contain updated ephemeris information for spacecraft position and orientation as updated by the bundle adjustment. Stored in ISIS standardized tables are spacecraft attitude (pointing), spacecraft position, target body rotation (orientation) and sun position. These data were extracted from each cube and ordered by image acquisition time.

The data stored in ISIS are instrument (OCAMS in this case) positions and rotations in the J2000 frame. Body positions and rotation quaternions in the ORX_OCAMS_MAPCAM or ORX_OCAMS_SAMCAM frames are transformed into the ORX_SPACECRAFT frame prior to applying interpolations. This is critical since both the OCAMS MapCam and SamCam instruments are used. In addition, these transforms may be dynamic in nature, particularly the struct offsets, which are fixed locations of selected spacecraft structures and science instruments. These may be updated during flight thus impacting frame origins. These struct offsets are applied by using the NAIF instrument id rather than the spacecraft id to determine observer location. This is achieved in ISIS for the OCAMS instruments by setting the *NaifSpkCode* to the OCAMS instrument frame id as indicated in Table 1. This is the case for both the OCAMS and OTES ephemeris data. For these types of transformations, the original kernels were loaded to determine these dynamic rotations at given acquisition times if needed. Once the interpolations are computed, transforms are again used to compute shift and rotation parameters from the OREX_SPACECRAFT frame into the ORX_OTES frame.

### Interpolation

For position data (spacecraft and sun), each XYZ component was fit using a cubic spline with time as the dependent variable. This provides interpolation models that are used to compute new XYZ position data for a given OTES observation at the time it was acquired.

For attitude data (spacecraft and target body) including velocity vectors, quaternion *spherical linear interpolation* ([SLERP](#)) fitting was utilized. This type of interpolation assumes rotation with uniform angular velocity. This is a reasonable assumption for this application as the spacecraft is in descent to the surface.

### Finding Surface Intercepts

Once in the ORX_OTES instrument frame, the boresight of the OTES instrument is retrieved from the OREX OTES instrument kernel and represents the look vector direction. The precise position of the OTES instrument is used as the observer position. Both of these data are transformed into the Bennu body-fixed frame (IAU_BENNU). Then, a ray along the look vector is traced to Bennu's surface using the NAIF toolkit DSK function [dskx02](#). The Bennu V021 5cm and 44cm shape models are prioritized, first checking for a surface intercept using the 5cm shape model. If it does not intersect the 5cm shape model, the 44cm global shape model is then used to determine the Bennu surface intercept location. This provides the XYZ surface point on Bennu of the OTES boresight from which latitude/longitude/radius coordinates are computed and updated for each observation.

## TAG Location

The TAG center on the V021 shape model is located at 55.91976219121° latitude, 41.877801555816° longitude. This coincides with the center of the TAGSAM sample collection head. These coordinates were computed from the OCAMS SamCam image 20201020T214950S101_sam_iofL2pan5_V013.fits at sample/line (511.41233, 477.783437) with a pixel resolution of 1.019 mm/pixel (Figure 7). The observation time of this image, which coincides with the TAG is 2020-10-20T21:49:50.0775798. This image was part of the control network and the ephemeris data has been updated by the bundle adjustment applied to the descent images.

## Results

We compared the differences in the V020 and V021 shape models by computing the surface intercept points of the latitudes and longitudes of all OTES observations on each of shape models. The spatial distance between the surface points was computed using the magnitude of the vector between the two surface points. The radius differences were computed by subtracting the radius (magnitude of the surface XYZ vectors) of the two intercept points. Virtually all the spatial offsets and radius differences were identical between shape models, indicating that they are in spatial agreement.

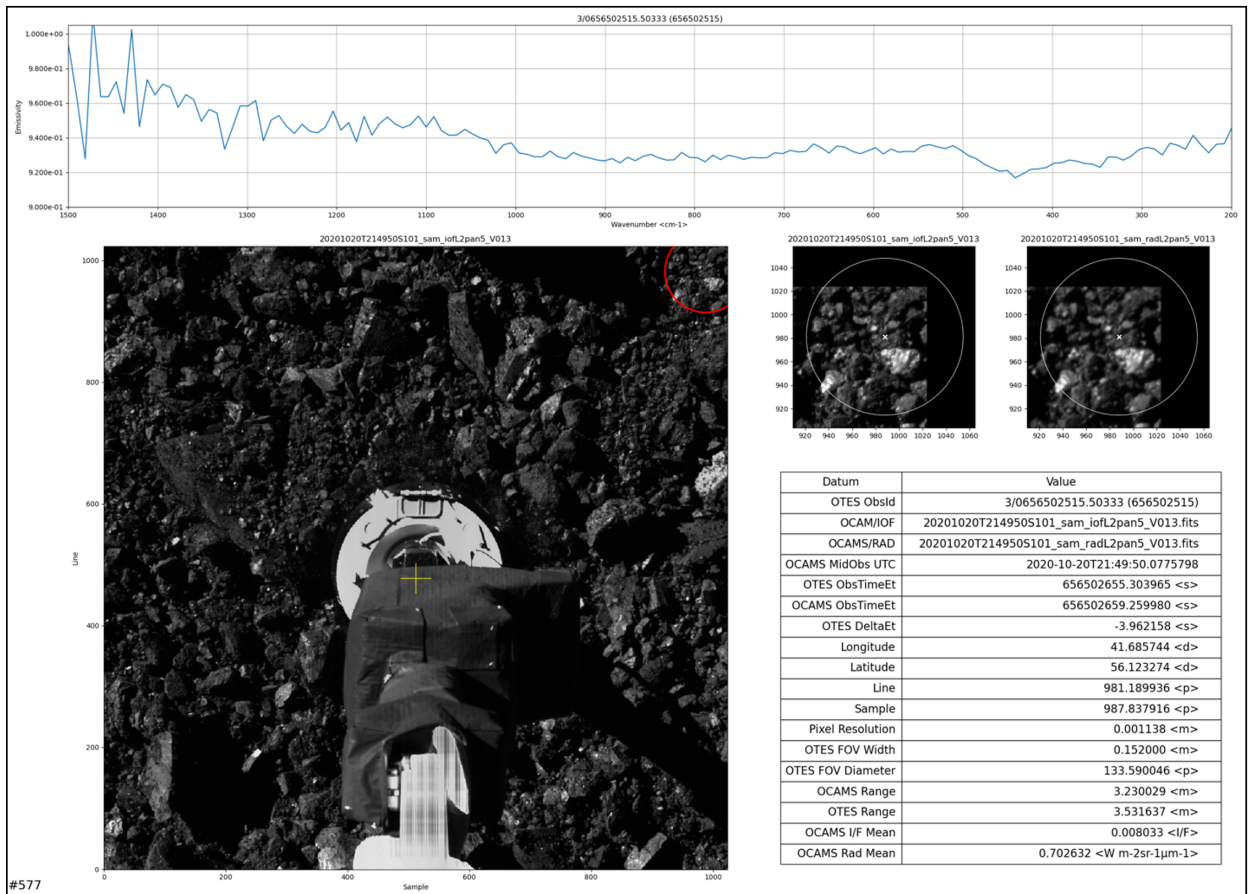

*Figure 7. The TAG location in SAMCAM image 20201020T214950S101_sam_iofL2pan5_V013.fits.*

With this knowledge, we then computed the a priori OTES latitude/longitude coordinate intercepts and the adjusted latitude/longitude using the OCAMS control interpolation data. The average spatial offset was ~37cm and the average radius difference was ~-0.87cm. Figure 8 shows the plot of these differences where the data points are color coded with time from TAG.

The OTES team reported that comparisons between the adjusted OTES latitude/longitudes computed using the controlled OCAM ephemeris were within 5 cm of the thermal model results.

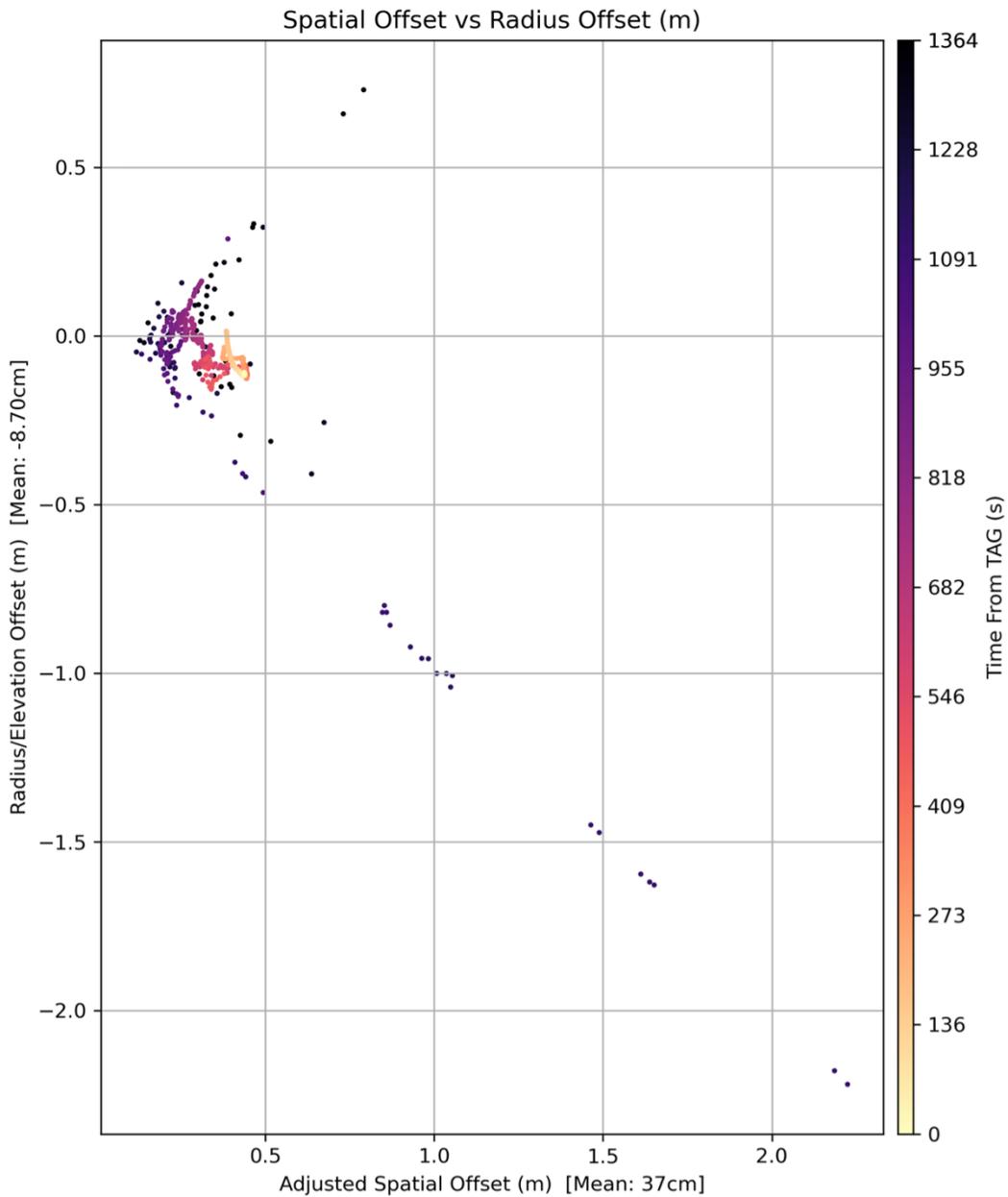

*Figure 8. Spatial and Radius offsets of adjusted OTES observation data using OCAMS control interpolation.*

# References

Acton, C.H., 1996. Ancillary Data Services of NASA's Navigation and Ancillary Information Facility. Planet. Space Sci., 44(1), 65-70, doi.org/10.1016/0032-0633(95)00107-7.

Becker, K.J., Berry K.L., Mapel, J.A., Walldren J.C., 2017. A New Approach to Create Image Control Networks in ISIS. *3rd Planetary Data Workshop and The Planetary Mappers Annual Meeting*, 12-15 June, Flagstaff, AZ, Abstract 7133.
https://www.hou.usra.edu/meetings/planetdata2017/pdf/7133.pdf

Becker, K.J., Becker, T.L., Edmundson, K.L., Golish, D.R., Bennett, C.A., Porter, N.A., DellaGiustina, D.N., Daly, M.G., Palmer, E., Weirich, J. , Barnouin, O.S., Philpott, L., Al Asad, M.M., Seabrook, J.A., Johnson, C.L., Rizk, B., Lauretta, D.S., 2020. Photogrammetric Control of Candidate Sample Sites on (101955) Bennu. *Lunar Planet Sci.*, *LI*, Abstract 2569.
https://www.hou.usra.edu/meetings/lpsc2020/pdf/2569.pdf

Bennett, C.A., DellaGiustina D.N., Becker, K.J., Becker T.L., Edmundson, K.L., Golish, D. R, Bennett, R. J., Burke, K.N., Cue, C.N.U., Clark, B.E., Contreras, J., Deshapriya, J.D.P., Drouet d'Aubigny, C., Jawin, E.R., Nolan, T.Q., Porter, N.A., Riehl, M.M., Roper, H.L., Rizk, B., Tang, Y., Zeszut, Z., Gaskell, R.W., Palmer E.E., Weirich, J.R., Al Asad, M.M., Philpott, L., Daly, M.G., Barnouin, O.S., Enos, H.L., Lauretta, D.S., 2020. A High-Resolution Global Basemap of (101955) Bennu. *Icarus*.
doi.org/10.1016/j.icarus.2020.113690

Brown, D.C., 1958. A solution to the general problem of multiple station analytical stereotriangulation. *RCA Data Reduction Technical Report No. 43*.
https://digital.hagley.org/08206139_solution#modal-close

Christensen, P.R., Hamilton, V.E., Mehall, G.L., Pelham, D., O'Donnell, W., Anwar, S., Bowles, H., Chase, S., Fahlgren, J., Farkas, Z., Fisher, T., James, O., Kubik, I., Lazbin, I., Miner, M., Rassas, M., Schulze, L., Shamordola, K., Tourville, T., West, G., Woodward, R., Lauretta, D., 2018. The OSIRIS-REx Thermal Emission Spectrometer (OTES) Instrument. Space Sci. Rev. 214 (87).
doi.org/10.1007/s112140-18-0513-6

DellaGiustina, D.N., Bennett C.A., Becker, K., Golish, D.R., Le Corre, L., Cook, D.A., Edmundson, K.L., Chojnacki, M., Sutton, S.S., Milazzo, M.P., Carcich, B., Nolan, M.C., Habib, N., Burke, K.N., Becker, T., Smith, P.H., Walsh, K.J., Getzandanner, K., Wibben, D.R., Leonard, J.M., Westermann, M.M., Polit, A.T., Kidd Jr, J.N., Hergenrother, C.W., Boynton, W.V., Backer, J., Sides, S., Mapel, J., Berry, K., Roper, H., Drouet d'Aubigny, C., Rizk, B., Crombie, M.K., Kinney-Spano, E.K., de León, J., Rizos, J.L., Licandero, J., Campins, H.C., Clark, B.E., Enos, H.L., Lauretta, D.S., 2018. Overcoming the Challenges Associated with Image-Based Mapping of Small Bodies in Preparation for the OSIRIS-REx Mission to (101955) Bennu. *Earth Space Sci.*, 5, 929-949.
doi.org/10.1029/2018EA000382

Edmundson, K.L., Cook, D.A., Thomas, O.H., Archinal, B.A., Kirk, R.L., 2012. Jigsaw: The ISIS3 Bundle Adjustment for Extraterrestrial Photogrammetry. *ISPRS Ann. Photogramm. Remote Sens. Spatial Inf. Sci.*, I-4, 203-208. doi.org/10.5194/isprsannals-I-4-203-2012

Edmundson, K.L., Becker, K.J., Becker, T.L., Bennett, C.A., DellaGiustina, D.N., Golish, D.R., Porter, N.A., Rizk, B., Drouet d'Aubigny, C., Daly, M.G., Palmer, E., Weirich, J., Barnouin, O.S., Philpott, L., Al Asad, M.M., Seabrook, J.A., Johnson, C.L., Lauretta, D.S., 2020. Photogrammetric Processing of OSIRIS-REx Images of Asteroid (101955) Bennu. *ISPRS Ann. Photogramm. Remote Sens. Spatial Inf. Sci.*, V-3-2020, 587-594. doi.org/10.5194/isprsannals-V-3-587-2020

Golish D.R., DellaGiustina, D.N., Clark, B.E., Bennett, C.A., Li, J.-Y., Zou, X.D., Lauretta, D.S., 2017. Photometric Modeling of Simulated Surface-Resolved Bennu Images. *3rd Planetary Data Workshop and the Planetary Mappers Annual Meeting*, 12-15 June, Flagstaff, AZ, Abstract 7069. https://www.hou.usra.edu/meetings/planetdata2017/pdf/7069.pdf

Rizk, B., Drouet d'Aubigny, C., Golish, D., Fellows, C., Merrill, C., Smith, P., Walker, M.S., Hendershot, J.E., Hancock, J., Bailey, S.H., DellaGiustina, D.N., Lauretta, D.S., Tanner, R., Williams, M., Harshman, K., Fitzgibbon, M., Verts, W., Chen, J., Connors, T., Hamara, D., Dowd, A., Lowman, A., Dubin, M., Burt, R., Whiteley, M., Watson, M., McMahon, T., Ward, M., Booher, D., Read, M., Williams, B., Hunten, M., Little, E., Saltzman, T., Alfred, D., O'Dougherty, S., Walthall, M., Kenagy, K., Peterson, S., Crowther, B., Perry, M.L., See, C., Selznick, S., Sauve, C., Beiser, M., Black, W., Pfisterer, R.N., Lancaster, A., Oliver, S., Oquest, C., Crowley, D., Morgan, C., Castle, C., Dominguez, R., Sullivan, M., 2018. OCAMS: The OSIRIS-REx Camera Suite. *Space Sci. Rev.*, 214, 26. doi.org/10.1007/s11214-017-0460-7

Sides, S.C., Becker, T.L., Becker, K.J., Edmundson, K.L., Backer, J.W., Wilson, T.J., Weller, L.A., Humphrey, I.R., Berry, K.L., Shepherd, M.R., Hahn, M.A., Rose, C.C., Rodriguez, K., Paquette, A.C., Mapel, J.A., Shinaman, J.R., Richie, J.O., 2017. The USGS Integrated Software for Imagers and Spectrometers (ISIS 3) Instrument Support, New Capabilities, and Releases. *Lunar Planet Sci., XLVIII,* Abstract 2739. https://www.hou.usra.edu/meetings/lpsc2017/pdf/2739.pdf